\documentclass[12pt,preprint]{emulateapj}

\def\am{\arcmin\hspace{-1.3mm}.\hspace{0.3mm}}
  
\def\P{\langle P \rangle} 
\def\tP{$\langle P \rangle$} 
\def\ETA{\langle \eta \rangle}

\slugcomment{Submitted to PASP}

\shorttitle{Dual-beam Optical Linear Polarimetry}
\shortauthors{Patat \& Romaniello}

\begin{document}


\title{Error Analysis for Dual-Beam Optical Linear 
Polarimetry}


\author{Ferdinando Patat \& Martino Romaniello\footnote{This paper is partially based on observations
made with ESO Telescopes at Paranal Observatory under programme IDs
066.A-0397, 69.C-0579, 069.D-0461 and 072.A-0025.}}

\affil{European Southern Observatory, K.Schwarzschildstr. 2, 
85748-Garching b. M\"unchen, Germany}
\email{fpatat@eso.org, mromanie@eso.org}




\begin{abstract} In this paper we present an error analysis for
polarimetric data obtained with dual-beam instruments. After recalling
the basic concepts, we introduce the analytical expressions for the
uncertainties of polarization degree and angle. These are then
compared with the results of Monte-Carlo simulations, which are also
used to briefly discuss the statistical bias.  Then we approach the
problem of background subtraction and the errors introduced by a
non-perfect Wollaston prism, flat-fielding and retarder plate
defects. We finally investigate the effects of instrumental
polarization and we propose a simple test to detect and characterize
it.  The application of this method to real VLT-FORS1 data has shown
the presence of a spurious polarization, which is of the order of
$\sim$1.5\% at the edges of the field of view. The cause of this
effect has been identified with the presence of rather curved lenses
in the collimator, combined with the non complete removal of
reflections by the coatings. This problem is probably common to all
focal-reducer instruments equipped with a polarimetric mode. An
additional spurious and asymmetric polarization field, whose cause
is still unclear, is visible in the $B$ band.
\end{abstract}




\section{Introduction}
\label{intro}

Performing polarimetry basically means measuring flux differences
along different electric field oscillation planes. In ground-based
astronomy this becomes a particularly difficult task, due to the
variable atmospheric conditions which make it difficult to detect the
relatively low polarization degrees which characterize most
astronomical sources \citep[a few percent; see for
example][]{leroy}. These fluctuations, in fact, introduce flux
variations among different polarization directions which can be
eventually mistaken for genuine polarization effects.

This problem has been solved in a number of different ways reviewed by
\citet{tinbergen} and to which we refer the reader for a detailed
description. In this paper we will focus on the so-called dual-beam
configuration, which is the most popular one for instruments
currently mounted at large telescopes. Despite new technologies, the
basic concept of astronomical dual-beam polarimeters \citep[see for
example][]{appenzeller,scarrot} has remained unchanged. A mask placed
on the focal plane, which prevents image (or spectra) overlap, is
followed by a Wollaston prism, which splits the incoming beam into two
rays characterized by orthogonal polarization states and separated by
a suitable angular throw. The rotation of polarization plane is
usually achieved with the introduction of a turnable retarder plate
(half or quarter wave for linear and circular polarization,
respectively) just before the Wollaston prism \citep[see for
example][]{schmidt}. Recently new solutions have been proposed, in
order to fully solve the problem in one single exposure (see Oliva
1997 and Pernechele et al. 2003 for an example application), but so
far they have been implemented in a few cases only.

Alternatives to Wollaston-based systems have been devised. They are
mainly based on the charge transfer in CCDs, which allows an {\it
on-chip} storage of two different polarization states, which are
obtained rotating a polarization modulator. After the pioneering work
of \cite{mclean83}, this technique, originally proposed by
P. Stockman, has been successfully applied in a number of instruments
\citep{mclean}.

In this work we address the most relevant problems which are connected
to two-beam polarimetric observations and data reduction.  The paper
is organized as follows. In Sec.~\ref{sec:basics} we introduce the
basic concepts of the problem and in Sec.~\ref{sec:analytical} we
recall the analytical expressions for the uncertainties of
polarization degree and angle, which are then compared to Monte-Carlo
simulations in Sec.~\ref{sec:mc}.  In the same section we also recap
the basics of polarization bias.  Sec.~\ref{sec:background} deals with
the effects of background on the polarization measurements and
Sec.~\ref{sec:flat} treats the flat-fielding
issues. Sec.~\ref{sec:wollaston} is devoted to the deviations of the
Wollaston prism from the ideal behaviour, the consequences of retarder
plate defects are addressed in Sec.~\ref{sec:HWP} while the effects of
post-analyzer optics are discussed in Sec.~\ref{sec:postwp}.
Sec.~\ref{sec:instr} is dedicated to the instrumental polarization and
Sec.~\ref{sec:fors1} deals with the case of VLT-FORS1. Finally, in
Sec.~\ref{sec:discussion} we discuss and summarize our results.

\section{Basic Concepts}
\label{sec:basics}

The polarization state of the incoming light can be described through
a Stokes vector $\vec{S}(I,Q,U,V)$ \cite[see, for
example,][]{chandra}. Its components, also known as Stokes parameters,
have the following meaning: $I$ is the intensity, $Q$ and $U$ describe
the linear polarization and $V$ is the circular polarization.  Linear
polarization degree $P$ and polarization angle $\chi$ are related to
the Stokes parameters as follows:

\begin{equation}
\label{eq:poldeg}
P=\frac{\sqrt{Q^2 + U^2}}{I} \equiv \sqrt{\bar{Q}^2 + \bar{U}^2}
\end{equation}

\begin{equation}
\label{eq:polang}
\chi=\frac{1}{2} \arctan \frac{U}{Q}
\end{equation}

where we have introduced the normalized  Stokes parameters $\bar{Q}=Q/I$ and 
$\bar{U}=U/I$. The above relations can be easily inverted to yield:

\begin{equation}
\label{eq:QU}
\bar{Q}=P \; \cos 2\chi \;\; ; \;\; \bar{U}=P \;\sin 2\chi
\end{equation}

Finally, the circular polarization degree, not discussed in this
paper, is simply $P_c=\bar{V}\equiv V/I$. For the sake of clarity, we
will set $V$=0 and neglect all circular polarization effects
throughout the paper.

The ideal measurement system for linear polarization is composed of a
half-wave retarder plate (HWP) followed by the analyzer, which is a
Wollaston prism (WP) producing two beams with orthogonal polarization
directions. In general, each of these elements can be treated as a
mathematical operator that acts on the input Stokes vector $\vec{S}$
\cite[see for example][]{shurcliff,goldstein}. 
What one measures on the detector is the intensities in the ordinary
and extraordinary beams at a given $HWP$ angle $\theta_i$, which are
related to the Stokes parameters by:

\begin{equation}
\label{eq:ord}
\left \{ 
\begin{array}{ccc}
f_{O,i} & = & 
\frac{1}{2} \left[ I + Q\;\cos 4\theta_i + U\;\sin 4\theta_i \right]\\
 & & \\
f_{E,i} & = & 
\frac{1}{2} \left[ I - Q\;\cos 4\theta_i - U\;\sin 4\theta _i\right]
\end{array}
\right.
\end{equation}

If the observations are carried out using $N$ positions for the HWP,
the whole problem of computing $I,Q$ and $U$ reduces to the solution
of the 2$N$ linear equations system given by Eqs.~\ref{eq:ord}. It is
clear that, having three unknowns ($I,Q$ and $U$), at least $N$=2 HWP
position angles have to be used.
 
Introducing the normalized flux differences $F_i$

\begin{equation}
\label{eq:F}
F_i \equiv \frac{f_{O,i} - f_{E,i}}{f_{O,i} + f_{E,i}} 
\end{equation}

and noting that $f_{O,i}+f_{E,i}=I$, Eqs.~\ref{eq:ord} reduce to the
following $N$ equations:

\begin{equation}
\label{eq:system}
F_i = \bar{Q} \cos 4\theta_i + \bar{U} \sin 4\theta_i =
P\;\cos(4\theta_i - 2\chi)
\end{equation}

We note that each $F$ parameter is totally determined by a single
observation, and it is therefore independent from sky conditions
changes. It is also worth mentioning that alternative approaches to
the normalized flux ratios exist. One example can be found in
\cite{miller}.

In principle, one can use any set of HWP angles to solve the problem,
but is is easy to show that adopting a constant step $\Delta
\theta$=$\pi/8$ is the optimal choice. In fact, besides minimizing the
errors of the Stokes parameters, this choice makes the solution of
Eqs.~\ref{eq:system} trivial:

\begin{equation}
\label{eq:Q}
\bar{Q}  =  \frac{2}{N} \; \sum_{i=0}^{N-1} F_i \cos(\frac{\pi}{2}i)
\end{equation}
\begin{equation}
\label{eq:U}
\bar{U}  = \frac{2}{N} \; \sum_{i=0}^{N-1} F_i \sin(\frac{\pi}{2}i)
\end{equation}

Finally, it prevents the ``power leakage'' \cite[see, for instance,][]{numrep}
when one is to perform a Fourier analysis (see below).

In the ideal case, the normalized flux differences $F_i$ obey
Eq.~\ref{eq:system}, which is a pure cosinusoid.  Since all possible
effects introduced by the HWP must reproduce after a full revolution,
it is natural to consider them as harmonics of a fundamental function,
whose period is $2\pi$.

Therefore, if $\theta_i=\pi i/8$, Eq.~\ref{eq:system} can be rewritten
as the following Fourier series:

\begin{displaymath}
F_i = a_0 + \; \sum_{i=1}^{N/2} 
a_k \; \sin(k \frac{2\pi i}{N}) + 
b_k \; \cos(k \frac{2\pi i}{N}) 
\end{displaymath}

where the Fourier coefficients are given by 

\begin{eqnarray}
\label{eq:fourcoeff}
a_0 & = & \frac{1}{N} \sum_{i=0}^{N-1} F_i \nonumber \\
a_k & = & \frac{2}{N} \; \sum_{i=0}^{N-1} F_i \; \cos(k\frac{2\pi i}{N}) 
\\
b_k & = & \frac{2}{N} \; \sum_{i=0}^{N-1} F_i \; \sin(k\frac{2\pi i}{N})
\nonumber
\end{eqnarray}

which are valid for $N=4,8,12,16$. Comparing Eqs.~\ref{eq:fourcoeff}
with Eq.~\ref{eq:system}, it is clear that the polarization signal is
carried by the $k=N/4$ harmonic. In a quasi-ideal case, all Fourier
coefficients are expected to be small compared with $a_{N/4}$ and
$b_{N/4}$ and deviations from this behaviour could arise from a number
of effects. For such an approach to the error analysis and for the
meaning of the various harmonics, the reader is referred to
\cite{fendt}.  Here we just notice that the $a_0$ term, which should
be rigorously null in the ideal case, is related to the deviations of
the WP from the ideal behaviour (see Sec.~\ref{sec:wollaston}).

In general, a Fourier analysis is meaningful when $N$=16, and can
reveal possible problems directly related to the HWP quality
(cf. Sec.~\ref{sec:HWP}).  In most cases, though, due to practical
reasons, one typically uses $N$=4 and, in that circumstance, a
different error treatment is required.

\section{Analytical Error analysis}
\label{sec:analytical}

Under the assumption that all relevant quantities are distributed
according to Gaussian laws, one can analytically derive simple
expressions for the corresponding errors of the final results.  As we
will see in the next section, this assumption is not always correct
and, when this happens, a numerical treatment is required in order to
test the analytical results and their range of validity.  Assuming
that the background level is the same in the ordinary and
extraordinary beams and that the read out noise can be neglected, the
analytical expression for the absolute error of $P$ can be readily
derived
\citep[see for example][]{miller} propagating the various errors
through the relevant equations\footnote{Here we consider photon shot
noise as the only source of random errors. Another potential source is
represented by the mis-positioning of the HWP with respect to the
optimal angles. However, as analytical solutions and numerical
simulations show, with the typical positioning accuracy nowadays
attainable ($<1^\circ$), the associated error of the polarization
degree and angle is negligible.}:

\begin{equation}
\label{eq:sigp}
\sigma_P = \frac{1}{\sqrt{N/2}\; SNR}
\end{equation}

where $SNR$ is the signal-to-noise ratio of the intensity image
($f_O+f_E$). The signal-to-noise one expects to achieve in the
polarization degree, $SNR_P=P/\sigma_P$, is simply given by:

\begin{displaymath}
SNR_P=\sqrt{\frac{N}{2}}\; P \; SNR
\end{displaymath}

As for the error of $\chi$, this is given by: 

\begin{equation}
\label{eq:sigchi}
\sigma_\chi = \frac{1}{2\sqrt{N/2} \; P \; SNR} \equiv \frac{\sigma_P}{2P}
\end{equation}

from which it is clear that, at variance with the polarization degree,
the accuracy of the polarization angle does depend on the intrinsic
polarization degree.

\section{Monte Carlo simulations}
\label{sec:mc}

The analytical treatment presented in Sec.~\ref{sec:analytical} relies
on the assumption that all relevant variables obey Gaussian
statistics. Numerical simulations are required in order to derive more
realistic distributions and to verify the validity of the analytical
results. One can easily implement the concepts we have developed until
now in a Monte Carlo (MC) code, which also allows higher
sophistication, like the inclusion of Poissonian noise. With this tool
one can readily investigate the effects of non-Gaussian distributions
of the derived quantities, the most important of which is the
systematic error of the polarization, as it was first pointed
out by
\cite{serkowski}.

\subsection{Linear Polarization Bias}
\label{sec:bias}

Due to the various noise sources, the vector components $\bar{Q}$ and
$\bar{U}$ are Normally distributed, but since $P$ is defined as the
quadrature sum of $\bar{Q}$ and $\bar{U}$, the statistical errors
always add in the positive direction, leading to a systematic increase
of the estimated polarization degree, thus introducing a bias. The
problem was addressed by several authors, both with analytical and
numerical methods
\citep{serkowski,wardle,simmons,clarke,sparks99}. We refer the reader
to those papers for a detailed description of the problem, while here
we just recall the basic concepts and we apply them to our case.

The polarization bias is usually quantified using a robust estimator,
supposedly giving a statistically significant representation of the
observed value, which is then compared with the input polarization in
order to derive the systematic correction. Different choices have been
adopted \cite[see][]{sparks99}.  Following the considerations by
\cite{wardle} we have here adopted the mode \tP\/
of the distribution in order to estimate the bias, that we therefore
define as $\Delta P = \P\/-P_0$, where $P_0$ is the input
polarization.  Once applied to to the observed data, the bias
correction $\Delta P$ tends to restore the symmetry of the deviation
distribution \cite[see for example][their Fig.~4]{sparks99}.

\begin{figure}
\plotone{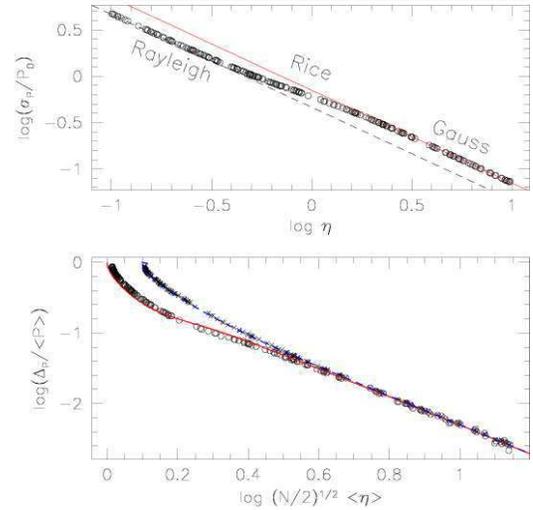}
\caption{\label{fig:deltap}Upper panel: comparison between the RMS error
on the polarization degree from MC simulations (circles) and
Eq.~\ref{eq:sigp}.  The dashed line traces the expected RMS error for
the Rayleigh distribution (see text).  Lower panel: bias estimated
using the mode (circles) and the average (crosses).  For comparison,
the solid curve traces the \cite{wardle} solution for the
mode, while the dashed line is the \cite{sparks99} solution for
the average.}
\end{figure}

In Fig.~\ref{fig:deltap} we show the results of our MC simulations for
the estimated RMS error of the polarization $\sigma_P$ (upper panel)
and polarization bias $\Delta P$ (lower panel) for $N$=4. Following
\cite{sparks99}, we have used $\eta\equiv P_0\cdot SNR$ and
$\ETA\equiv \P\cdot SNR$ as independent variables in our plots.  As we
had anticipated, $\sigma_P$ follows the analytical prediction of
Eq.~\ref{eq:sigp} when $\eta>$2.  For lower values of $\eta$,
$\sigma_P$ tends to be systematically smaller than the analytical
prediction and it converges to the value expected for the Rayleigh
distribution (dashed line),which becomes a very good approximation for
$\eta<$0.5, provided that $SNR>$3. For intermediate values of $\eta$,
the distribution is described by a Rice function \citep{rice}. In
conclusion, one can safely use the analytical solution given by
Eq.~\ref{eq:sigp} for $\eta\geq$2 only.

As for the polarization bias, we have plotted it as a function of
measurable quantities, namely the signal-to-noise and the {\it
observed} polarization $\P$.

\begin{figure}
\plotone{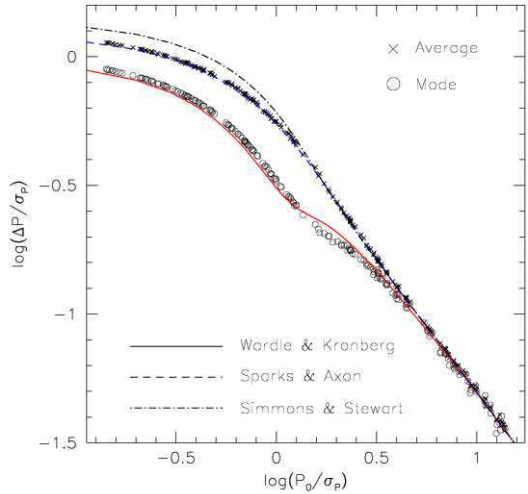}
\caption{\label{fig:bias} Comparison between systematic bias $\Delta P$ 
and random error $\sigma_P$ for different estimators as a function of
input polarization signal-to-noise ratio.}
\end{figure}

The results of our MC simulations, as plotted in the lower panel of
Fig.~\ref{fig:deltap}, are in good agreement with the analytical
solution found by \cite{wardle} for the same statistical
estimator. For comparison, we have also plotted the results one
obtains when the average is adopted (crosses). For
$\sqrt{N/2}\;\ETA>$4 the relation between $\log(\Delta P/\P\/)$ and
$\log(\sqrt{N/2}\;\ETA)$ is well approximated by a linear law. A least
squares fit gives the following result:

\begin{displaymath}
P_0=\P\/ \left [ 1- \left( \frac{0.62}{\sqrt{N/2}\;\ETA}\right)^{1.92} \right ]
\end{displaymath}

which can be used to correct the observed polarization values
according to the input signal-to-noise ratio, measured polarization
and number of HWP positions. In general, the bias effect is present
even at reasonably high values of $SNR$ when the polarization is small
and $\sigma_P$ and $\Delta P$ tend to become similar, so that the
systematic bias correction is comparable to the random uncertainty of
the polarization.  This is better seen in Fig.~\ref{fig:bias}, where
we have plotted the ratio $\Delta P$ over $\sigma_P$ as a function of
$P_0/\sigma_P\equiv SNR_P$ deduced from our simulations.  As
anticipated, for low values of $SNR_P$, the ratio between $\Delta P$
and $\sigma_P$ tends to unity, with some variations among different
estimators. For $SNR_P\geq$3 the bias correction is less than 10\% of
the expected accuracy, and it is, therefore, negligible. Moreover,
above that threshold, all estimators give practically identical
results.

In Fig.~\ref{fig:bias} we have plotted, for comparison, the function
computed by \cite{simmons}, who have used a Maximum
Likelihood estimator in order to evaluate the bias. As these authors have
shown, this is the best estimator for $SNR_P\leq$0.7, while for $SNR_P>$0.7
the mode, first used by \cite{wardle}, should be used.

\section{The Effects of the Background}
\label{sec:background}

Until now we have assumed that one is able to perfectly subtract the background
contribution. This is most likely the case when one is to perform polarimetric
measurements on point-like sources, since in that situation local background
subtraction is in most cases straightforward.

We remark that the background, whatever its nature is, must be
subtracted before the calculation of normalized Stokes parameters
\cite[see also][]{tinbergen}, so that possible background polarization
is {\it vectorially} removed.

If we assume that the object is characterized by $P_o$ and $\chi_o$
and the background by $P_b$ and $\chi_b$, the two polarization fields
can be expressed using the Stokes vectors defined as $\vec{S_o}(I_o,
I_o P_o \cos 2\chi_o, I_o P_o \sin 2\chi_o)$ and $\vec{S_b}(I_b, I_b
P_b \cos 2\chi_b, I_b P_b \sin 2\chi_b)$, where we have neglected any
circular polarization. Since Stokes vectors are additive \cite[see for
example][]{chandra}, the resulting polarization field is described by
$\vec{S}=\vec{S_o}+\vec{S_b}$ and, therefore, the total polarization
is given by the following formula:

\begin{displaymath}
P =\frac{I_o P_o}{I_o+I_b} \sqrt{1+ r^2 + 2r\cos [2(\chi_o-\chi_b)]}
\end{displaymath}

where $r=(I_b P_b)/(I_o P_o)$, i.e. the ratio between the polarized fluxes of
background and object. The corresponding polarization angle is

\begin{displaymath}
\chi=\frac{1}{2}
\arctan{\frac{\sin 2\chi_o + r\;\sin 2\chi_b}
{\cos 2\chi_o + r\;cos 2\chi_b}}
\end{displaymath}

Clearly, the background is going to influence significantly the object
when $r\gtrsim1$. For $r\sim1$ one can write:

\begin{displaymath}
P\simeq \frac{P_o}{\sqrt{2}} \sqrt{1+\cos [2(\chi_o-\chi_b)]}
\end{displaymath}

which implies that, for comparable polarized fluxes, the resulting
polarization is nulled when the polarization fields are {\it perpendicular}
($|\chi_o-\chi_b|=\pi/2$).

\section{Flat fielding}
\label{sec:flat}

One of the basic problems one has to face when reducing the data
produced by dual-beam instruments is the flat-fielding. Due to the
fact that image splitting occurs after the focal mask, the collimator
and the HWP, one would in principle need to obtain flat exposures with
all optical components in the light path.  Unfortunately, in all
practical conditions, this introduces strong artificial effects due to
the strong polarization typical of flat-field sources (either twilight
sky or internal screens). In principle one can reduce this effect
using the continuous rotation of the HWP as a depolarizer. This is
implemented, for example, in EFOSC2, currently mounted at the ESO-3.6m
telescope \citep{efosc} and it is effective only if the HWP rotation
time is much shorter than the required exposure time.  The
depolarizing effect can also be achieved by averaging flats taken the
same set of HWP angles used for the scientific exposures. In fact,
with the usage of the optimal angle set (see Sec.~\ref{sec:basics}),
one has that

\begin{displaymath}
\sum_{i=0}^{N-1} f_{O,i} =\frac{N}{2} I
\end{displaymath}

and a similar expression for $f_{E,i}$, which do not contain any
polarization information.  The problem is that this is true only if
the source is stable in intensity, which is surely not the case for
the twilight sky and probably not really true for most lamps.

An alternative solution (at least for imaging) is the usage of a set
of twilight flats obtained without HWP and WP. If on the one hand this
eliminates source polarization, on the other hand it does not allow
for a proper flat field correction. In fact, while the pixel-to-pixel
variations are properly taken into account, the large scale patterns
are not, due to the beam split that maps a given focal plane area into
two distinct regions of the post-WP optics and detector. Moreover,
these calibrations do not carry any information about possible spatial
effects introduced by the HWP and the WP. However, as the simulations
show, this problem becomes milder if some redundancy is
introduced. For example, if one uses $N$=4 HWP positions, the ordinary
and extraordinary ray will just swap when the angle differs by $\pi/4$
within each of the two redundant pairs ($f^i_O=-f^{i+2}_E$, see
Eqs.~\ref{eq:ord}). This tends to cancel out the flat-field effect and
becomes more efficient if the maximum redundancy ($N$=16) is used.
It must be noticed, however, that time dependent effects, like fringing,
may affect the redundant pairs in a different way, therefore decreasing
the cancellation efficiency.

\section{Effects of a non-ideal Wollaston prism}
\label{sec:wollaston}

So far we have assumed that our system is described by
Eqs.~\ref{eq:ord}, i.e. that the Wollaston prism splits incoming
unpolarized light into identical fractions. A deviation from this ideal
behaviour can be described by the introduction of a new parameter $t$
in Eqs.~\ref{eq:ord}, which can be reformulated as follows:

\begin{equation}
\label{eq:WP}
\left \{ 
\begin{array}{ccl}
f_{O,i} & = & t \left[ I + Q\;\cos 4\theta_i + U\;\sin 4\theta_i \right]\\
 & & \\
f_{E,i} & = & (1-t) \left[ I - Q\;\cos 4\theta_i - U\;\sin 4\theta _i\right]
\end{array}
\right.
\end{equation}

An ideal system is obtained for $t=\frac{1}{2}$. Now, for unpolarized light
($Q$=$U$=0), these new equations give $f_{O,i}=tI$ and $f_{E,i}=(1-t)I$ for
all HWP angles, so that all normalized flux differences turn out to be
identical, i.e. $F_i=2t-1$. Therefore, the value of $t$ can be directly 
estimated observing an unpolarized source. 

In the simplest situation, where $N$=2, neglecting the presence of the
$t$ term would lead to a spurious polarization degree
$P=\sqrt{2}\;(2t-1)$, with a polarization angle $\chi=\pi/8$. It is
interesting to note that this is not the case, for example, when
$N$=4. In that situation, in fact, because all redundant $F$'s are
identical, Eqs.~\ref{eq:Q} and \ref{eq:U} would correctly yield null
Stokes parameters.

The problem when the incoming light is polarized is more complicated,
since the normalized flux differences are not anymore a linear
combination of $\bar{Q}$ and $\bar{U}$, as one can verify from
Eqs.~\ref{eq:WP}:

\begin{equation}
\label{eq:kappa}
F_i=\frac{\kappa + \bar{Q}\cos 4\theta_i + \bar{U} \sin 4\theta_i}
{1 + \kappa \bar{Q}\cos 4\theta_i + \kappa \bar{U} \sin 4\theta_i}
\end{equation}

where we have set $\kappa =2t-1$ ($|\kappa|\leq$1, $\kappa$=0 in the
ideal case). If $\kappa$ is known, one can correct the observed $f_O$
and $f_E$ dividing them by $2t$ and $2(1-t)$ respectively before
following the procedure adopted for the ideal case. Of course, if the
source has a known polarization (e.g. a polarized standard), one can
use this information together with the observed $F$ ratios to derive
$\kappa$ for each HWP angle, according to the following relation:

\begin{displaymath}
\kappa_i=\frac{\bar{Q}\cos 4\theta_i + \bar{U} \sin 4\theta_i - F_i}
{F_i \; (\bar{Q} \cos 4\theta_i + \bar{U} \sin 4\theta_i)}
\end{displaymath}

If the input polarization is unknown, then one can in principle derive
$\kappa$ from the observations themselves, provided that $N\geq$4. In
fact, after introducing the parameter

\begin{equation}
\label{eq:g}
g_j = \frac{1+F_{j-1} \; F_{j+1}}{F_{j-1} + F_{j+1}}
\end{equation}

and using Eqs.~\ref{eq:kappa}, it is easy to demonstrate that

\begin{equation}
\label{eq:kappa2}
\kappa_j = g_j \pm \sqrt{g^2_j-1}
\end{equation}

where $j=1,...,N/2-2$ and the positive sign refers to the case
$g_j\leq 0$. For instance, for $N=4$, one can determine two
independent estimates of $\kappa$, which can be averaged to improve on
the accuracy.

It is interesting to note that, when $P\ll 1$, Eq.~\ref{eq:kappa} can
be approximated as $F_i\simeq \kappa + \bar{Q}\cos 4\theta_i + \bar{U}
\sin 4\theta_i$. If $N$ is a multiple of 4 (i.e. if the $F$ function
is sampled for an integer number of periods $\pi/2$), one has that

\begin{equation}
\label{eq:a0}
\kappa \approx \frac{1}{N}\;\sum_{i=0}^{N-1} F_i \equiv a_0
\end{equation}

which clarifies the meaning of the $a_0$ term in the Fourier series
(see Eq.~\ref{eq:fourcoeff}).

It is worth mentioning that the redundancy on the $F$ parameters does
eliminate the effects of a non ideal WP to a large extent. For
example, a blind application of Eqs.~\ref{eq:Q} and
\ref{eq:U} to the case of $N$=4 gives the following result:

\begin{displaymath}
\frac{2}{N} \; \sum_{i=1}^{N-1}  F_i \cos(4\theta_i) = 
\bar{Q} \; \frac{1-\kappa^2}{1-\kappa^2 \bar{Q}^2}
\end{displaymath}

and a similar expression for $\bar{U}$. If the polarization is small
($P_0 \leq 0.1$), we have that $\kappa^2 \bar{Q}^2\ll1$ and therefore
the application of the ideal case procedure to a non ideal situation
would lead to a value of $\bar{Q}$ which is $(1-\kappa^2)$ times
smaller than the real one. Since the same is true for $\bar{U}$, the
resulting polarization $P$ will also be $(1-\kappa^2)$ times smaller
than the input value, while the polarization angle remains unchanged.
For example, if $|\kappa|\leq 0.2$, $\sigma_P/P$ is less than 4\%.

Eqs.~\ref{eq:WP} describe a particular case only, where the incoming
unpolarized flux is distributed into two fractions $t$ and $(1-t)$.
More in general, one should replace the term $(1-t)$ with an
independent parameter $s$, so that the fraction of light split by the
WP in the ordinary and extraordinary ray become uncorrelated. Using
the same procedure, it is easy to demonstrate that one can estimate
the ratio $\kappa_j=(t-s)/(t+s)$ still using Eq.~\ref{eq:kappa2}.

\begin{figure}
\plotone{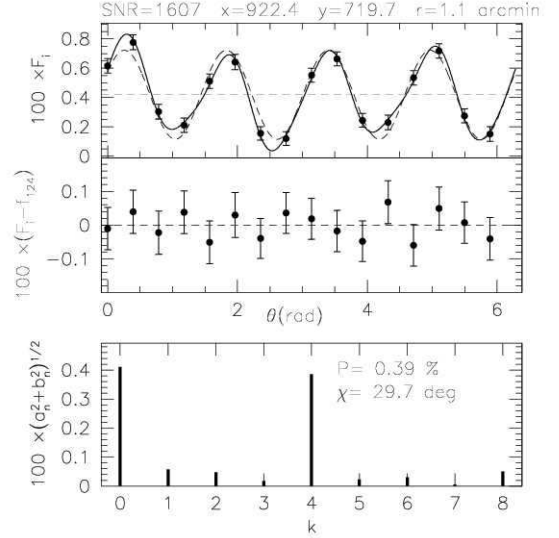}
\caption{\label{fig:fourier}Example of Fourier Analysis applied to archival 
VLT-FORS1 observations of a bright star in the $V$ passband (see
text). Upper panel: normalized flux differences. Partial
reconstructions using 8 harmonics (solid curve) and the 4-th harmonic
only (dashed curve) are traced. The dashed horizontal line is placed
at the average of $F$ values ($a_0$). Central panel: residuals after
subtracting the $k$=1,2,4 components. Lower panel: harmonics power
spectrum.}
\end{figure}

Another effect we have investigated is the possibility that the
difference in polarization direction between the ordinary and the
extraordinary ray of the WP is not $\pi/2$. If we call $\Delta\alpha$
the deviation from this ideal angle, using the general expression of
the Mueller matrix for a linear polarizer \citep[see for
example][]{goldstein} and deriving the expressions for the normalized
flux ratios, one gets:

\begin{displaymath}
F_i\simeq \frac{\bar{Q}}{2} [ \cos(4\theta_i-2\Delta\alpha) + \cos 4\theta_i] +
\frac{\bar{U}}{2} [ \sin(4\theta_i-2\Delta\alpha) + \sin 4\theta_i]
\end{displaymath}

where the approximation is valid under the assumption that
$P\ll1$. With the aid of this expression it is easy to conclude that
for reasonably small values of $\Delta\alpha$ ($\leq$10$^\circ$), the
implied errors are of the order of 0.05\% on the polarization degree
and 5$^\circ$ on the polarization angle, irrespective of the number of
HWP positions used.

\section{HWP defects}
\label{sec:HWP}

In the ideal case, the normalized flux differences are modulated by
the HWP rotation according to Eq.~\ref{eq:system}, which is a pure
cosinusoid. If defects are present on the HWP, like dirt or
inhomogeneously distributed dust, one can expect spurious flux
modulations, which are not related to the polarization of the incoming
light and can reduce the performance of the instrument. As a
consequence, error estimates based on pure photon statistics are
systematically smaller than the actual ones.

These kinds of problems can be investigated with the aid of Fourier
Analysis, following the procedure we have outlined at the end of
Sec.~\ref{sec:basics}.  This method becomes particularly effective
when the observations are taken sampling the full HWP angle range,
i.e. 2$\pi$ which, given the choice of the optimal angle set
$\theta_i$=$\pi i/8$, implies $N$=16 retarder plate positions. In
these circumstances, one is able to determine the Fourier coefficients
$a_k$ and $b_k$ for the first 8 harmonics, the fundamental ($k=1$)
being related to local transparency fluctuations which repeat
themselves after a full revolution, like dirt or dust.  By definition
(cf. Eq.~\ref{eq:system}), the fourth harmonic is directly related to
the linear polarization, i.e.  $a_4\equiv\bar{Q}$ and
$b_4\equiv\bar{U}$. All other harmonics, with the only remarkable
exception of the second one ($k$=2), are simply overtones of harmonics
with lower frequencies and include part of the noise generated by the
photon statistics, which is present at all frequencies and is
therefore indicated as white noise. For this reason, the global random
error is often estimated as the signal carried by the harmonics having
$k$=3,5-8 \cite[see for example][their Appendix~A.3]{fendt}.

The second harmonic deserves a separate discussion. Ideally, the HWP
operates as a pure rotator of the input Stokes vector, with the
advantage that one does not need to rotate the whole instrument in
order to analyze different polarization planes. In the real case,
being the HWP usually constructed using bi-refringent materials, it is
affected by the so-called pleochroism.  This is a wavelength dependent
variation of the transmission that takes place when the direction of
the incoming light is changed with respect to the crystal lattice.
Due to the way the HWP is manufactured, the crystals have an axial symmetry,
which gives a period of $\pi$. Therefore, this effect is seen as the 
$k$=2 component.

In Fig.~\ref{fig:fourier} we show a real case, where we have applied
this analysis to archival data obtained with the FOcal Reducer/low
dispersion Spectrograph (hereafter FORS1), which is currently mounted
at the Cassegrain focus of ESO-VLT 8.2m telescope
\citep{szeifert}. 

A bright (and supposedly unpolarized) star was observed using $N$=16
HWP positions. First of all, the Fourier Analysis indicates the
presence of a small deviation of the WP from the ideal behaviour
($\kappa\approx a_0\simeq$4.1$\times$10$^{-3}$, see
Sec.~\ref{sec:wollaston}).  Then, a clear polarization is detected, at
the level of about 0.4\% while all other components are smaller than
0.05\% (this polarization is actually an instrumental effect present
in FORS1. See Sec.~\ref{sec:instr}).  The effective significance
of harmonics other than $k$=4 can be judged on the basis of the
expected errors of the Fourier coefficients. For example, using the
expression of $a_k$ one finds that

\begin{displaymath}
\sigma_{a_k}= \frac{2}{N\; SNR}\; \sqrt{\sum^N_{i=1} 
\cos^2 k \frac{2\pi i}{N}}
\end{displaymath}

With this kind of analysis, one can see that in the example of
Fig.~\ref{fig:fourier}, $a_k$ and $b_k$ are consistent with a null
value for $k$=3,5-7 (see the central panel). As for the $k$=1,2 harmonics, 
the Fourier coefficients are non null at a 2$\sigma$ level. Since for the test
star it was $SNR\sim$1600, it is clear that to detect $k$=1,2 harmonics of this
amplitude (0.05\%), a $SNR\geq$3000 is required.

It is important to notice that the presence of these non-null
harmonics is implicitly corrected for when one has a sufficient number
of HWP positions covering the maximum period 2$\pi$. In the most
common case, where $N$=4 angles spaced by $\pi$/8 are used, one can
derive the fundamental (i.e. linear polarization, period $\pi$/2) and
the first overtone (period $\pi$/4) only. The latter corresponds to
the $k$=8 component of the $N$=16 cases, which therefore carries the
high frequency information only. As a consequence, if other harmonics
are present, they are not properly removed and contribute to the final
error, practically setting the maximum accuracy one can achieve,
irrespective of the SNR.  Numerical simulations performed assuming a
virtually infinite SNR show that in the presence of $k=1$ and $k=2$
components, the usage of $N$=4 HWP angles leads to systematic errors
which are of the same order of the amplitude of the two
harmonics. From this and the example reported in
Fig.~\ref{fig:fourier}, one can estimate that the absolute maximum
accuracy reachable with FORS1 using $N$=4 is of the order of 0.05\%.

Another typical problem which affects the retarder plates is the
chromatic dependence of the angle zero point. This is usually measured
by means of a Glan-Thompson prism and it can change by more than
5$^\circ$ across the optical wavelength range. The computed
polarization angle can be corrected simply adding the HWP angle offset
for the relevant wavelength (or effective wavelength in the case of
broad band imaging). See, for example, \cite{szeifert}.

Finally, we have investigated the effects produced by a deviation
$\Delta\beta$ from the nominal phase retardance of an HWP ($\pi$). Using
the general expression of the Mueller matrix for a retarder 
\citep[see for example][]{goldstein}, the normalized flux ratios turn out 
to be

\begin{eqnarray}
\nonumber
F_i & = &\bar{Q} [\cos 4\theta_i -\sin^2 2\theta_i
(1-\cos\Delta\beta)] +\\
\nonumber
    & + &\frac{1}{2}\bar{U}\sin 4\theta_i \; (1+\cos\Delta\beta) +
    \bar{V}\sin 2\theta_i \sin\Delta\beta
\end{eqnarray}

from which it is clear that the measured linear polarization depends
on the circular polarization of the input signal. For $V$=0 and
$\Delta\beta \leq$10$^\circ$ the corresponding absolute error of the
polarization degree is less than 0.05\%, while the outcome on the
polarization angle is negligible. For $V\neq$0 the exact effect
depends on the ratio between circular and linear polarization
degrees. For example, for $\Delta=$10$^\circ$ and $Q=U=V=$0.01, the
absolute error of the computed polarization degree is about 0.1\% for
$N$=4, which decreases to 0.01\% for $N$=16. It is worthwhile noting
that this defect would be detected by a Fourier analysis as a
component with a period $\pi$ and whose intensity is
$|\bar{V}\sin\Delta|$.

\section{Effects of post-analyzer optics}
\label{sec:postwp}

Typically the analyzer is followed by additional optics, like filters,
grisms and camera lenses which, due to their possible tilt with
respect to the optical axis, may behave as poor linear polarizers. In
the most probable case where the polarization is produced by
transmission (see also next section), the properties of post-analyzer (PA)
components can be described by the following approximate Mueller
matrix:

\begin{eqnarray}
\nonumber
M_{PA}\simeq A
\left (
\begin{array}{ccccccc}
1   & & Bc & & Bs  & & 0\\
Bc  & & 1  & & 0   & & 0\\
Bs  & & 0  & & C   & & 0\\
0   & & 0  & & 0   & & C\\
\end{array}
\right) & &
\end{eqnarray}

\begin{figure}
\plotone{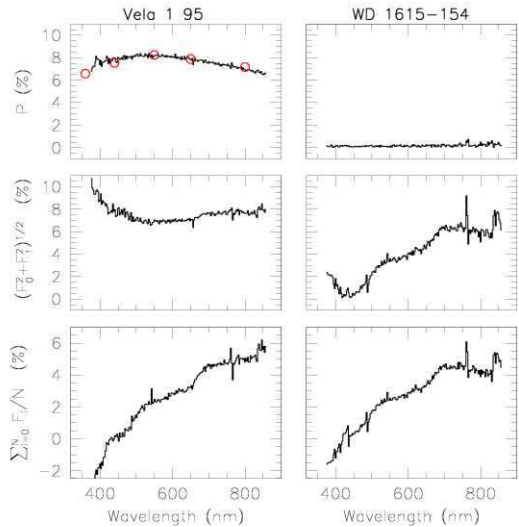}
\caption{\label{fig:postanal} VLT-FORS1 observations of Vela~1 95 
(left panels) and WD~1615-154 (right panels). The plots show the
linear polarization derived with $N$=4 HWP positions (upper panels),
$N$=2 (middle panels) and $\sum_{i=0}^N F_i/N$ for $N$=4 (lower
panels. See text for more details).  The original spectra were
obtained with the 300V grism and a slit of 1$^{\prime\prime}$; for
presentation they have been binned to 25 \AA.  The empty circles in
the upper left panel mark the broad-band polarimetric measurements for
Vela~1 95 (UBVRI, from left to right).}
\end{figure}

\vspace{2mm}

where $A\approx$1, $C\approx$1, $c=\cos 2\varphi$, $s=\sin 2\varphi$
and $\varphi$ is the polarization angle (which can change across the
field of view) while $B$ is related to the polarization degree
introduced by the PA optics. This expression can be deduced from the
general formulation
\citep[see][Eq. 4.63]{keller} after applying the usual matrix rotation
\citep[see for example][Eq. 2.5]{keller}.
If $\vec{S_0}=(I_0,Q_0,U_0,V_0)$ is the input Stokes vector, the
effect of PA optics can be evaluated computing the Stokes vectors that
correspond to the ordinary and extraordinary beams produced by the WP,
transforming them using the operator $M_{PA}$ and using the resulting
intensity components to compute the normalized flux differences
$F_i$. After simple calculations, one arrives at the following
expression:

\begin{displaymath}
F_i\simeq B\cos 2\varphi + \bar{Q_0}\cos 4\theta_i + \bar{U_0} \sin\theta_i
\end{displaymath}

were we have assumed that $|B|\ll 1$, i.e. that the linear
polarization induced by the PA optics is small. Given this expression,
it is clear that the redundancy in the HWP positions ($N$=4, 8, 16)
eliminates this problem, since the additive term $B\cos 2\varphi$ is
not modulated by the HWP rotation, while for $N$=2 the derived
polarization degree and angle would be affected, possibly in a severe
way.  If the optimal HWP angle set has been used, it is easy to verify
that $B\cos 2\varphi = \sum_{i=0}^{N-1} F_i/N$, which is identical to
Eq.~\ref{eq:a0}. This means that, in a first approximation, it is not
possible to disentangle between an imperfect WP and the presence of
polarization in the PA optics.  Therefore, the fact that
$a_0\simeq$0.4\% in Fig.~\ref{fig:fourier}, can actually be attributed
to both kinds of problems. The PA optics effect becomes definitely
stronger when these include highly tilted components, like the
grisms. This is very well illustrated by the two examples of
Fig.~\ref{fig:postanal}, where we show the results obtained using
VLT-FORS1 archival data of a highly polarized star (Vela~1 95,
$\alpha$=09:06:00, $\delta$=$-$47:19:00) and an unpolarized star
(WD~1615-154, $\alpha$=16:17:55, $\delta$=$-$15:35:51)\footnote{See 
\tt http://www.eso.org/instruments/fors/inst/pola.html}, which were observed
on the optical axis, where the instrumental polarization is known to be null 
\citep{szeifert}.

In both cases the polarization degree deduced using $N$=2 (central
panels) is markedly different from that derived with $N$=4 (upper
panels) and the deviation is particularly severe for the unpolarized
object. As one can finally notice, the resulting values of $B\cos
2\varphi$ show a strong wavelength dependency and are higher than 5\%
at about 800 nm.  It is interesting to note that $B\cos 2\varphi\sim$0
at about 450 nm, i.e. at the wavelength where the anti-reflection
coatings are optimized (see next section). This fact, together with
the marked wavelength dependency and the much lower level seen in
broad band imaging (see Fig.~\ref{fig:fourier}), strongly suggest that
the effect seen in Fig.~\ref{fig:postanal} is indeed produced by the
tilted surfaces of the grism.

\section{Effects of instrumental polarization}
\label{sec:instr}

So far, we have assumed that all optics preceding the analyzer do not
introduce any polarization. Of course, this is not generally true  
\cite[see for example][and Leroy 2000 for a
general introduction to the subject]{tinbergen}.

To show the effect of instrumental polarization, we assume that the
pre-analyzer optics, which include telescope mirrors, collimator, HWP
and so on, introduce an artificial polarization, which depends on the
position in the field. For the sake of simplicity, we assume these
optics act as a non-perfect linear polarizer, characterized by a
position dependent polarization degree $p(x,y)$ and polarization angle
$\varphi(x,y)$. This can be described by the following Mueller matrix:

\begin{eqnarray}
\nonumber
M_I(x,y)=\frac{1}{1+p}
\left (
\begin{array}{ccccc}
1   & & pc     & & ps     \\
pc  & & 1-ps^2 & & psc    \\
ps  & & pcs    & & 1-pc^2 \\
\end{array}
\right) & &
\end{eqnarray}

where we have neglected circular polarization and we have set $s=\sin
2\varphi$ and $c=\cos 2\varphi$. For $p$=0 one obtains a totally
transparent component, while $p$=1 gives an ideal linear polarizer.

If $\vec{S_0}(I_0,Q_0,U_0)$ is the Stokes vector describing the input
polarization state, it will be transformed by pre-analyzer optics into the
vector $\vec{S_1}= M_I \cdot \vec{S_0}$ before entering the analyzer:

\begin{eqnarray}
\label{eq:instpol}
\begin{array}{llcl}
I_1 &(1+p) & = & I_0 + pc Q_0 + ps U_0\\
Q_1 &(1+p) & = & Q_0 + pc I_0 -  ps^2 Q_0 + pcs U_0\\
U_1 &(1+p) & = & U_0 + ps I_0 +  pcs Q_0 - pc^2 U_0 
\end{array}
\end{eqnarray}

and, therefore, the measurements would lead to $\vec{S_1}$ which would then
need to be corrected for the instrumental effect inverting 
Eqs.~\ref{eq:instpol}, provided that $p(x,y)$ and $\varphi(x,y)$ are known.

Of course, if the observed source is known to be unpolarized, $p$ and
$\varphi$ can be derived immediately, for example by placing a single
target on different positions of the field of view or observing an
unpolarized stellar field.

If the source is polarized, the problem becomes much more complicated,
since Eqs.~\ref{eq:instpol} are non linear in $c$ and $s$. The
solution can be simplified assuming that $p\ll1$ and $P_0\ll1$, which
is a reasonable hypothesis in most real cases, since instrumental
polarization is typically less than a few percent. In these
circumstances, Eqs.~\ref{eq:instpol} can be rewritten as follows:

\begin{eqnarray}
\label{eq:approx}
\begin{array}{lll}
I_1 & \simeq & I_0 \\
Q_1 & \simeq & Q_0 + p I_0 \;\cos 2\varphi\\
U_1 & \simeq & U_0 + p I_0 \;\sin 2\varphi
\end{array}
\end{eqnarray}

It is important to note that the instrumental polarization is not
removed by the local background subtraction. Moreover, it is
independent of the object's intensity; in fact, using the previous
expressions one can verify that

\begin{displaymath}
P=\sqrt{P_0^2 + p^2 + 2P_0p \; \cos [2(\chi_0-\varphi)]}
\end{displaymath}

where $P_0$ and $\chi_0$ are the input polarization degree and
angle. From this expression it is clear that when $P_0\gg p$ it is
also $P\approx P_0$, while in the case that object and instrumental
polarization are comparable ($p\approx P_0)$, the observed polarization
is approximately given by

\begin{displaymath}
P\simeq \sqrt{2} P_0 \sqrt{1+\cos [2(\chi_0-\varphi)]}
\end{displaymath}

which, according to the value of $(\chi_0-\varphi)$, gives values that
range from 0 to 2$P_0$. It is important to notice that the main
difference between instrumental polarization and a polarized
background is that the latter is effective only when $BGR\gtrsim I$
(see Sec.~\ref{sec:background}), while the former acts regardless of
the object intensity and what counts is its polarization.

\begin{figure}
\plotone{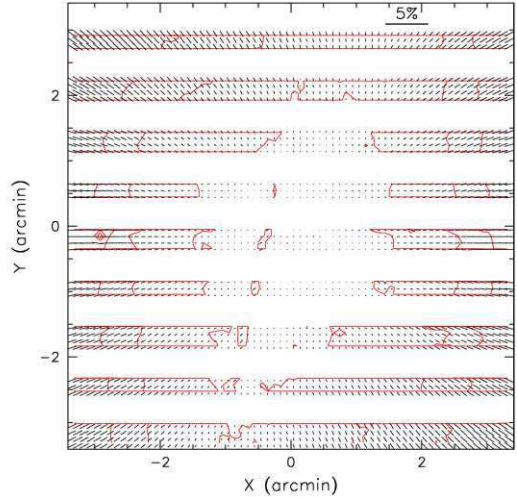}
\caption{\label{fig:insmapB} FORS1 instrumental polarization map in the $B$
band. The contours trace 0.3\%, 0.6\% and 0.9\% polarization levels.
Coordinates, expressed in arcminutes, refer to the geometrical center
of the detector.}
\end{figure}

\begin{figure}
\plotone{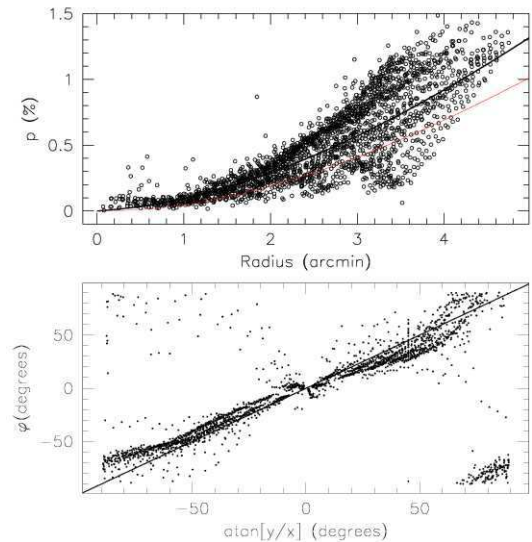}
\caption{\label{fig:radprofB} Upper panel: FORS1 instrumental polarization 
radial profile for the $B$ band. Each point is the result of a
30$\times$30 px binning in the original images. Radius, expressed
in arcminutes, is computed from the geometrical center of the
detector. The thick line traces a linear least squares fit, while the
thin line is the polarimetric ray-tracing prediction.  Lower panel:
instrumental polarization angle as a function of pixel polar
angle. The solid line is not a fit to the data, but rather has unit
slope and zero intercept.}
\end{figure}

With the aid of these approximate expressions (Eqs.~\ref{eq:approx}),
one can easily evaluate the instrumental polarization, provided that
the input polarization field is known and the observed source covers a
large fraction of the instrument field of view. In fact, solving
Eqs.~\ref{eq:approx} for $\varphi$ and $p$ yields:

\begin{displaymath}
\tan 2\varphi \simeq \frac{U_1-U_0}{Q_1-Q_0}
\end{displaymath}

and

\begin{displaymath}
p_1\simeq\frac{Q_1-Q_0}{I_0 \cos 2\varphi} \;\;\;\;; \;\;\;\;
p_2\simeq\frac{U_1-U_0}{I_0 \sin 2\varphi}
\end{displaymath}

where $p_1$ and $p_2$ are two independent estimates of $p$, that can
be averaged to increase the accuracy.

As it is well known, the night sky shows a polarization which varies
according to the ecliptic and galactic coordinates \cite[see][for an
extensive review]{leinert}. It is mostly dominated by the zodiacal
light polarization, which reaches its minimum, below a few percent, at
the anti solar position
\citep{roach}. Since this is not expected to vary on the scales of a
few arcminutes, in principle, relatively empty fields represent
suitable targets for panoramic polarization tests, provided that the
signal-to-noise ratio per spatial resolution element is of the order of
several thousands.

\section{The case of FORS1 at ESO-VLT}
\label{sec:fors1}

To show an example application, we have performed a test using real
data obtained with FORS1 at the ESO-VLT. In this instrument, the
polarimetric mode is achieved inserting in the beam a super-achromatic
HWP and a WP, which has a throw of about 22$^{\prime\prime}$
\citep{szeifert}.

We have identified in the ESO archive three sets of data obtained in
rather empty fields in $B$, $V$ and $I$ passbands. Equatorial
coordinates ($\alpha$, $\delta$), ecliptic longitude and latitude
($\lambda$, $\beta$), helio-ecliptic longitude
($\lambda-\lambda_\odot$), sky polarization degree and angle
($P_{sky}$, $\chi_{sky}$) are reported in Table~\ref{tab:empty} for
the different fields. In all three cases the signal-to-noise ratio
achieved on the sky background $I_{sky}$ in the combined images is
larger than $SNR\simeq$200 per pixel. With such a signal and for a
typical 1\% polarization, the bias effect is expected to be small (see
Fig.~\ref{fig:deltap}) and the RMS error of the polarization degree,
according to Eq.~\ref{eq:sigp}, is of the order of 0.3\%, while the
uncertainty of $\chi$ is about 9$^\circ$ (see Eq.~\ref{eq:sigchi}). In
order to further increase the accuracy and to allow for outlier
rejection, we have computed a clipped average in 30$\times$30 px bins
which, given the FORS1 detector scale (0$^{\prime\prime}$.2
px$^{-1}$), translates into an angular resolution of
6$^{\prime\prime}$.

\begin{table}
\caption{\label{tab:empty} Data for the empty fields}
\begin{tabular}{lccccccc}
Filter & $\alpha$ & $\delta$    & $\lambda$ & $\beta$ & $\lambda-\lambda_\odot$ & $P_{sky}$ & $\chi_{sky}$ \\
       & hh:mm:ss & dd:mm:ss    & $^\circ$  & $^\circ$& $^\circ$                & \%        & $^\circ$\\
\hline
B      & 03:32:17 & $-$27:44:24 & $-$41.1  & $-$45.1 & 128.1 & 6.45 & +68.4 \\
V      & 13:58:03 & $-$31:22:21 & 218.6    & $-$18.1 & $-$159.7 & 1.74 & $-$53.8 \\
I      & 20:36:08 & $-$13:06:39 & 308.0    & +5.3    & $-$110.0 & 1.34 & +68.7 \\
\end{tabular}
\end{table}

\begin{figure}
\plotone{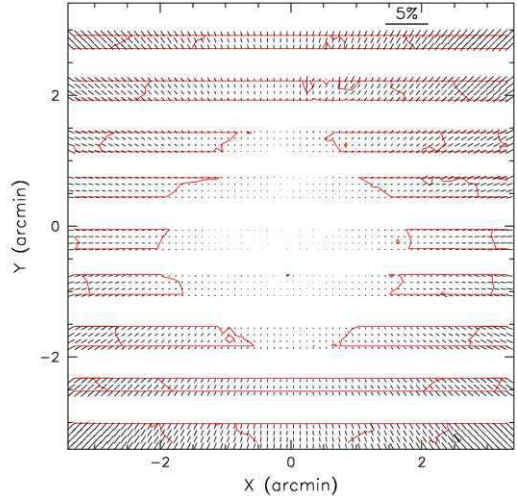}
\caption{\label{fig:insmapV} Same as Fig.~\ref{fig:insmapB} for the $V$ band.
The dark segment marked by a circle in the lower right corner
indicates the values obtained from an unpolarized standard star.}
\end{figure}

Since the instrumental polarization on the optical axis, measured with
unpolarized standard stars, is smaller than 0.03\% \citep{szeifert},
one can be confident that the sky background polarization field
($P_{sky}, \chi_{sky}$) measured close to that area is not affected by
spurious effects (the values are reported in the last two columns of
Table~\ref{tab:empty}). Therefore, we can easily compute $p(x,y)$ and
$\varphi(x,y)$ using the method previously outlined, where
$I_0$=$I_{sky}$, $Q_0=I_{sky} P_{sky} \cos 2\chi_{sky}$ and
$U_0=I_{sky} P_{sky} \sin 2\chi_{sky}$. The results of these
calculations are presented in Figs.~\ref{fig:insmapB},
\ref{fig:insmapV} and \ref{fig:insmapI}. With the remarkable
exception of the $B$ band, the instrumental polarization of FORS1
shows a quasi-symmetric radial pattern. For example, for the $V$
filter, the instrumental polarization remains below 0.1\% within 1
arcmin from the geometrical center of the detector, while it grows up
to $\sim$0.6\% at 3 arcmin, to reach the maximum, i.e. $\sim$1.4\%, at
the corners of the field of view.

\begin{figure}
\plotone{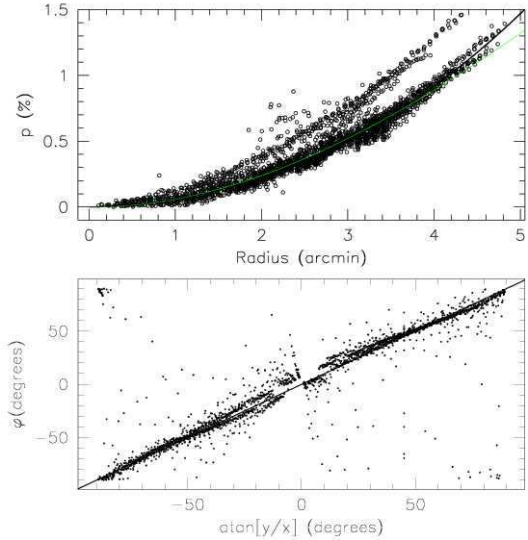}
\caption{\label{fig:radprofV} Same as Fig.~\ref{fig:radprofB} for the 
$V$ band.}
\end{figure}

\begin{figure}
\plotone{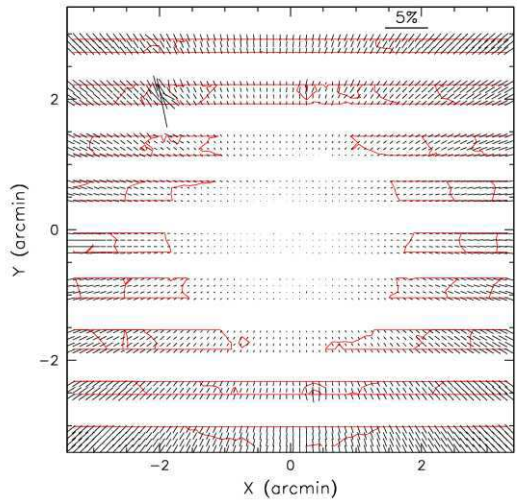}
\caption{\label{fig:insmapI}Same as Fig.~\ref{fig:insmapB} for the $I$ band.}
\end{figure}

\begin{figure}
\plotone{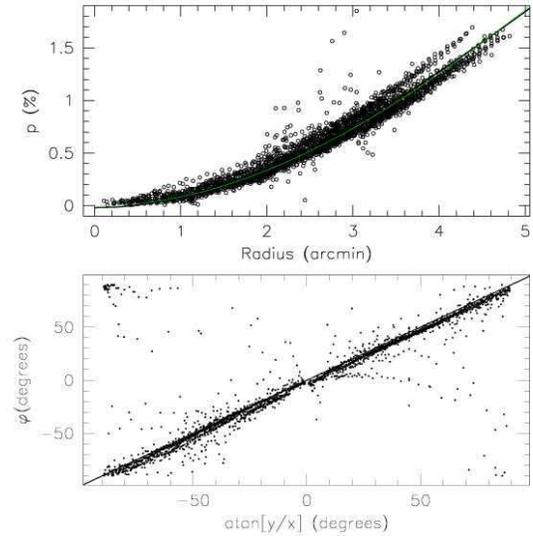}
\caption{\label{fig:radprofI}Same as Fig.~\ref{fig:radprofB} for the $I$ band.}
\end{figure}

This is illustrated in the upper panels of Figs.~\ref{fig:radprofB},
\ref{fig:radprofV} and \ref{fig:radprofI},
where we have plotted the estimated instrumental polarization for each
30 $\times$ 30 px bin as a function of its average distance $r$ from
the center. The deviation from a perfect central symmetry is
distinctly shown by the dispersion of the points, which is larger than
the measurement error. Particularly marked is the case of the $B$
band, which shows a strong azimuthal dependence and deserves a
separate discussion (see next section). For the $V$ band there is a
systematic deviation from central symmetry for a polar angle
$\alpha=\arctan(y/x)$ between 10$^\circ$ and 80$^\circ$.  This region
is probably disturbed by the presence of a saturated star and a
reflection caused by the HWP, visible in the input images. Excluding
these points, a linear least squares fit to the $V$ data gives

\begin{displaymath}
p(r) = 0.012 r + 0.046 r^2 +0.002 r^3
\end{displaymath}

where $r$ is expressed in arcminutes (Fig.~\ref{fig:radprofV}, solid
curve).  The $I$ band shows the smoothest behaviour and the
observations are described very well by the following polynomial:

\begin{displaymath}
p(r) = -0.017 r + 0.105 r^2 -0.006 r^3
\end{displaymath}

The absolute RMS deviations shown by the data from the best fits are
of the order of 0.05\%, so that in these two passbands the spurious
polarization can be corrected with an accuracy which is comparable to
that dictated by the photon statistics. In both cases, but especially
in the $I$ band, the pattern is remarkably radial, as shown in the
lower panels of Figs.~\ref{fig:radprofV} and \ref{fig:radprofI}, where
we have plotted $\varphi$ as a function of polar angle $\alpha$.

In order to verify these results for the $V$ filter, we have carried
out a test observing an unpolarized standard star placed in the lower
right corner of the detector. Measured polarization was
$P$=0.92$\pm$0.04\% and $\chi$=$-$48$^\circ \pm$1$^\circ$.4 while,
according to the previous analysis, the expected instrumental
polarization in that position is $p$=0.96\% and
$\varphi=-$51$^\circ$.9, which are in very good agreement with each
other (see also Fig.~\ref{fig:insmapV}, lower right corner).

Once the instrumental polarization is mapped, one can correct for it
using the approximate Eqs. \ref{eq:approx}, which hold when $p$ and
$P_0$ are small and only if the instrumental polarization is produced
by a linear polarizer preceding the analyzer.

\subsection{The cause of instrumental polarization in FORS1}
\label{sec:cause}

As we have seen, the spurious polarization detected in FORS1 in $V$
and $I$ passbands has a clear central symmetry, it is null on the
optical axis \cite{szeifert}, shows a radial pattern and grows with
the distance from the optical axis. All these facts suggest that this
must be generated by the optics which precede the analyzer,
i.e. within the collimator. In fact, when a light beam enters an
optical interface along a non-normal direction, the component of the
transmitted beam perpendicular to the plane of incidence is
attenuated, according to Fresnel equations \citep[see for
example][]{born}. As a consequence, the emerging beam is linearly
polarized in a direction which is parallel to the plane of
incidence. If the surface is curved, as it is the case of lenses, the
incidence angle quickly increases moving away from the optical axis
and this, in turn, produces an increase in the induced polarization.
The effect becomes more pronounced if the lens is strongly curved,
i.e. if the curvature radius is comparable to its diameter. Of course,
on the optical axis, the incidence angle is null, so that no
polarization is produced. Therefore, at least from a qualitative point
of view, the polarization induced by transmission has all the required
features to explain the observed pattern.

The polarization induced by transmission can be easily evaluated using
the appropriate expression for the corresponding Mueller matrix 
\citep[see][Eq. 4.63]{keller}. For a typical refraction index $n$=1.5 and 
an incidence angle of 30$^\circ$, refraction through an uncoated glass
would produce a polarization of about $B$=1.7 \% {\it per optical
surface}. This polarization is usually reduced in a drastic way
(i.e. down to 0.1-0.2\%) by anti-reflection (AR)
coatings. Nevertheless, since the effect of multiple surfaces is
roughly additive, in the presence of numerous and pretty curved
lenses, residual polarization can be non negligible.  Another
important aspect is that this mechanism has no effect on circular
polarization, in agreement with the fact that no instrumental circular
polarization has been measured in FORS1 \citep{bagnulo}. 

We have run polarization ray-tracing simulations, including telescope
mirrors, collimator lenses and AR coatings. This kind of calculation
allows one to describe in detail the optical system, taking into
account partial polarization cancellation produced by symmetries
within the optical beams and the depolarizing effect of AR coatings.

The standard resolution collimator of FORS1 contains 3 lenses and a
doublet, all treated with a single-layer MgF$_2$ quarter-wave AR
coating at 450 nm \citep{seifert}. The ray-tracing calculations
\citep{avila} show that indeed the polarization induced by
transmission is not totally removed by the AR coatings. For $V$ and
$I$ filters, a best fit to the simulated data gives a radial
dependence which is very similar to the results we have derived from
the experimental data. The deviation is maximum at the edges of the
field of view, where the ray-tracing model gives a polarization which
is $\sim$0.08\% and $\sim$0.05\% smaller than what is actually
observed in $V$ and $I$ respectively (see also
Figs.~\ref{fig:radprofV} and ~\ref{fig:radprofI}, upper panel, thin
curves). Possible explanations for this small discrepancy are to be
identified with imperfections in the AR coatings and with the effects of
non-orthogonal incidence on the HWP.

In principle, since single-layer AR coatings are optimized for one
specific wavelength (450 nm in the case of FORS1), the residual
polarization is expected to be higher at other
wavelengths. Simulations have been run in order to sample the
wavelength range 400-900 nm and they show that the expected wavelength
dependency can be very well approximated by the following linear
relation:

\begin{displaymath}
\frac{p(\lambda)}{p(550)} \simeq 0.02 + 1.73\times10^{-3} \; \lambda
\end{displaymath}

where $\lambda$ is expressed in nm. This relation predicts pretty
accurately the $\sim$40\% relative increase we indeed see passing from
$V$ to the $I$ passband and can, thus, be safely used to predict the
effect in the $R$ band.

\begin{figure}
\plotone{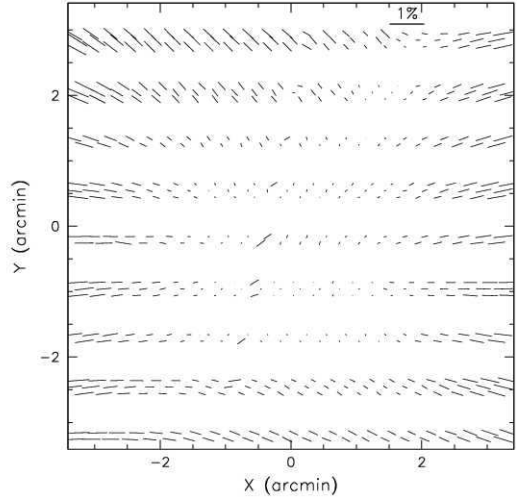}
\caption{\label{fig:residB}Residual field obtained subtracting the 
ray-tracing model from the observed polarization in the $B$ band.}
\end{figure}

According to the simulations, one would expect that the spurious
polarization in $B$ is about 25\% smaller than in $V$. But, as we have
already mentioned, this passband shows a rather weird behaviour and
does not conform to the model predictions.  In fact, the polarization
pattern strongly deviates from central symmetry, displaying a marked
azimuthal dependence (Fig.~\ref{fig:insmapB}, left panel).  This
becomes more evident looking at the radial profile presented in
Fig.~\ref{fig:radprofB} (upper panel): the purely radial dependence is
clearly disturbed by an asymmetric field. In some directions the
polarization field grows much faster than in others, producing a great
spread in the observed data and, in most of the cases, the observed
polarization is larger than what is predicted by the polarization
ray-tracing (thin solid curve). The deviations from a centrally
symmetric pattern reach up to 0.5\% (see Fig.~\ref{fig:residB}),
making the correction in the $B$ band quite difficult, and certainly
not feasible using simple smooth functions as in the cases of $V$ and
$I$. Rather, a much more accurate correction can be obtained by
interpolating the map of Fig.~\ref{fig:insmapB} at the required field
position. We must remark, however, that a rigorous correction for
this secondary effect will be possible only once its physical reason
is identified and its mathematical description is formulated.

We have tried to reproduce the observed behaviour introducing defects
in the system, like a weak linear polarization from the HWP and the
presence of linear polarization in the post analyzer optics. In both
cases the effect is completely different from what we see in the $B$
band. Therefore, the physical reason of this phenomenon is still
unclear (see also the discussion in the next section). What we can say
here is that the deviation from central symmetry is present, though to
a much smaller extent, also in the $V$ band. This is shown by the
contours at constant polarization, which are clearly box-shaped (see
Fig.~\ref{fig:insmapV}), while in the $I$ band they are practically
circular (see Fig.~\ref{fig:insmapI}). The conclusion is that this
additional effect, whatever its origin is, becomes more severe at
shorter wavelengths.

We must notice that our method heavily relies on the assumptions that
the instrumental polarization is null on the optical axis and the sky
background polarization is constant across the field of view. As a
matter of fact, the night sky polarization is not very well studied. The
only extensive analysis we could find in the literature is the one by
\cite{wolstencroft}, who concluded that the polarization structure varies
in scale from a few degrees to about 30$^\circ$, i.e. on scales which
are much larger than the field of view of FORS1 (6\am8 $\times$
6\am8). In order to explain the deviations we observe from a centrally
symmetric pattern, one would need a variation in the sky polarization
of the order 0.5\% on a scale of a few arcmin. Even though this seems
to be quite a large gradient, in principle we cannot exclude it. Only
further tests will clarify the nature of the effect we see in the $B$
band.

\section{Discussion and Conclusions}
\label{sec:discussion}

Dual-beam polarizers coupled to 2D arrays provide a tool to perform
panoramic imaging polarimetry and multi-object spectro-polarimetry.
In these instruments, the atmospheric fluctuations problem is solved
by obtaining simultaneous measurements of two orthogonal polarization
states. Of course, the use of a WP has also some drawbacks, as the
flat-fielding issue discussed in Sec.~\ref{sec:flat}.  With the only
exception of this feature, data reduction and analysis are totally
similar to other polarimetric systems, as we have shown with both
analytical and numerical approaches (Secs.~\ref{sec:analytical} and
\ref{sec:mc}). 

When the targets to be studied are extended and cover a large fraction
of the field-of-view, accurate background subtraction becomes an
issue, whose effects we have investigated in
Sec.~\ref{sec:background}. This is particularly important when the
background is not the simple sky background, but it has a complicated
structure. This is the case, for instance, for a faint supernova
projected onto a galactic spiral arm.

Another problem that may reduce the performance of a dual-beam
polarimeter is the imperfect behaviour of the WP. In
Sec.~\ref{sec:wollaston} we have discussed this issue and presented a
test to determine possible deviations from the ideal case.  As an
example, we have applied it to the FORS1 archive data we have
described in Sec.~\ref{sec:instr}. Using an object-free region roughly
in the center of the field-of-view we have used Eqs.~\ref{eq:g} and
\ref{eq:kappa2} to compute $t$ (see Eq.~\ref{eq:WP}), which turns out
to be $t$=0.502$\pm$0.001, i.e. perfectly compatible with the value
derived from the Fourier analysis (Sec.~\ref{sec:HWP}).  As we have
shown, the redundancy introduced by having $N\geq 4$ strongly reduces
this problem, even in the cases where $t$ differs by about 10\% from
the ideal case ($t$=0.5). This is the case also for the presence of
linear polarization in the post-analyzer optics, whose effects are
practically eliminated by the redundancy (Sec.~\ref{sec:postwp}).

Finally, we have addressed the instrumental polarization issue,
described its consequences and proposed an easy test to detect any
spurious effect with rather high accuracy (Sec.~\ref{sec:instr}).  As
an example, we have applied it to archival FORS1 data and we have
detected an instrumental polarization pattern, roughly centrally
symmetric (for $V$ and $I$) and with a radial dependency. The presence
of this spurious polarization affects all objects placed at distances
larger than 1\am5 from the optical axis with intrinsic polarizations
of a few percent or less. The problem becomes particularly severe when
$p\simeq P$: in that case the measured Stokes parameters can be
severely wrong.  For objects filling most of the field of view, there
always will be regions affected by this problem.  Moreover, the
correct sky background estimate, which is absolutely necessary to
recover the intrinsic object field in the outer parts of the galaxy,
becomes impossible, if the instrumental polarization is not taken care
of properly. The spurious field must be removed before one is able to
estimate the background contribution.  Both our data and ray-tracing
simulations show that the effect is wavelength dependent. In the case
of FORS1, a strong deviation from central symmetry is seen in the $B$
band and we have interpreted this as a signature of an additional
effect, yet to be explained, not included in the ray-tracing
simulations that, in contrast, reproduce quite well the observed data
in $V$ and $I$. 

One possible source of asymmetric instrumental polarization is the
unrelieved stress birefringence in the optical glasses, due to thermal
strain and mechanical loading \citep[see for
example][]{theocaris}. This phenomenon is known to introduce a
retardance which, in turn, can change the polarization status of
incoming polarized light (the effect is null if the light is
unpolarized).  Since the incoming radiation is certainly polarized by
FORS1 in a differential way across the field of view, this
would also imply that the secondary effect should be weaker where the
centrally symmetric component is smaller. The fact that this is indeed
the case (see Fig.~\ref{fig:residB}) and also that the retardance is
expected to grow faster than $\lambda^{-1}$ seem to suggest that this
is a plausible explanation for the asymmetric component. If this is
indeed the case, then it is not possible to correct the measured
linear polarization just subtracting vectorially the residual field
(like the one shown in Fig.~\ref{fig:residB}) simply because the
effect of retardance depends on the polarization state of the incoming
light. This requires a more sophisticated treatment, necessarily based
on the exact knowledge of the physical mechanism and its mathematical
description through Mueller matrices formalism.

In general, instrumental polarization induced by transmission is most
likely common to all focal reducers equipped with a polarimetric mode.
While the overall pattern should be a general feature of these
instruments, the exact radial dependence may change according to the
optical design and the curvature of the lenses. The method we have
described in this paper allows an accurate way of characterizing the
instrument and a tool to correct for this effect.

\begin{acknowledgements}
This paper is partially based on observations made with ESO Telescopes
at Paranal Observatory under programme IDs 066.A-0397, 69.C-0579,
069.D-0461 and 072.A-0025.  The authors would like to thank T.
Szeifert for his kind support and collaboration, J.Walsh and
S. Bagnulo for interesting discussions, G. Ruprecht and
W. Seifert for providing us with the optical design specifications of
FORS1, R. Tommasini for introducing us to polarimetric ray-tracing,
S. D'Odorico and H. Dekker for their kind support and G. Avila for his
polarimetric ray-tracing calculations. Special thanks go to C. Keller,
for his invaluable advices, clarifications and help during the
analysis of the instrumental polarization of FORS1. Finally, we would
like to thank an anonymous referee for her/his comments and suggestions,
which helped a lot to improve the quality of the paper.
\end{acknowledgements}

\end{document}